\documentclass[twocolumn,showpacs,fleqn,nobibnotes]{revtex4-1}
\usepackage{float,subfig,graphicx,epsfig,epsf}
\usepackage{multirow}
\usepackage{url}
\usepackage{amsmath,units}
\usepackage[usenames,dvipsnames,svgnames,table]{xcolor}
\usepackage[colorlinks=true,linkcolor=red,urlcolor=blue,citecolor=blue]{hyperref}
\usepackage{comment}

\begin{document}

	\thispagestyle{empty}

\title{Exclusive Photoproduction $J/\psi$ in Peripheral Pb-Pb}
\author{M. B. GAY DUCATI\footnote{beatriz.gay@ufrgs.br} AND S. MARTINS\footnote{sony.martins@ufrgs.br}}

\affiliation{High Energy Physics Phenomenology Group, GFPAE  IF-UFRGS \\
Caixa Postal 15051, CEP 91501-970, Porto Alegre, RS, Brazil}

\begin{abstract}
The exclusive photoproduction of the $J/\psi$ state is investigated in peripheral AA collisions for the energies available at the LHC, $\sqrt{s}=2.76$ TeV and $\sqrt{s}=5.02$ TeV. In order to evaluate the robustness of the light-cone color dipole formalism, previously tested in the ultraperipheral regime, the rapidity distribution and the nuclear modification factor ($R_{AA}$) were calculated for three centrality classes: 30\%-50\%, 50\%-70\% and 70\%-90\%. In the peripheral regime, three scenarios were considered. In the scenario 1, a similar formalism adopted in the UPC regime is used; in the scenario 2, one considers that only the spectators in the target are the ones that interact coherently with the photon; in the scenario 3, the photonuclear cross section is modified using the same geometrical constraints applyed in the scenario 2. The results obtained from the three scenarios were compared with the ALICE measurements (only $J/\psi$ at the moment), showing a better agreement in the more complete approach (scenario 3), mainly in the more central regions (30\%-50\% and 50\%-70\%) where the incertainty is smaller.
\end{abstract}

 
  \maketitle
  
\section{Introduction}

The ALICE collaboration measured the $J/\psi$ hadroproduction in peripheral collisions Pb-Pb, with $\sqrt{s}=2.76$ TeV, revealing an excess in the production of the meson in very small transverse momentum ($p_T<0.3$ GeV/c) in the range rapidity $2.5<y<4.0$ \cite{PRL116-222301}. This excess was quantified by the nuclear modification factor, $R_{AA}$, which reaches the values 7 (2) in the centrality classes 70\%-90\% (50\%-70\%) within the rapidity range $2.5<y<4.0$. Considering the $J/\psi$ excess could be generated from exclusive photoproduction, the nuclear photoproduction of the $J/\psi$ was calculated in peripheral collisions from the factorization adopted in the ultraperipheral collisions (UPC), with cross section separated in two factors: the equivalent photon flux, $N\left(\omega, b\right)$, and the photonuclear cross section, $\sigma_{\gamma A\rightarrow VA}$. In our approach, the peripheral regime is analysed considering three scenarios: in the scenario 1 is applyed a photon flux with b-dependence and a more realistic electromagnetic form factor is used; in the scenario 2 a geometrical cut is applyed in the photon flux ensuring that only the spectators in the target will interact coherently with the photon; in scenario 3, for completeness, the restriction adopted in the scenario 2 is extended on the photonuclear cross section. Using these three scenarios, the rapidity distribution and the nuclear modification factor, $R_{AA}$, were estimated for 30\%-50\%, 50\%-70\% and 70\%-90\% centrality classes.

To calculate the $R_{AA}$, it was adopted the expression developed in \cite{PLB734-314}, 
\begin{eqnarray}
\scalebox{0.7}{$R_{AA}^{hJ/\psi}=\dfrac{N_{AA}^{J/\psi}}{BR_{J/\psi\rightarrow l^+l^-}\cdot N_{events}\cdot\left(A\times\varepsilon\right)_{AA}^{J/\psi}\cdot\left\langle T_{AA}\right\rangle \cdot\sigma_{pp}^{hJ/\psi}}$}\label{eq:raa},
\end{eqnarray}
where the measured number of $J/\psi$ ($N_{AA}^{J/\psi}$) is corrected for acceptance and efficiency $\left(\mathcal{A}\times\varepsilon\right)_{AA}^{J/\psi}\sim 11.31\%$ and branching ratio $\text{BR}_{J/\psi\rightarrow \mu^+\mu^-}=5.96$\%. Then, the result is normalized to the equivalent number of MB events ($N_{events}\simeq 10.6\times 10^7$), defined in \cite{PLB734-314},  average nuclear overlap function $\left(\langle T_{AA}\rangle\right)$, calculated from \cite{PRC88-044909}, and proton-proton inclusive $J/\psi$ production cross section ($\sigma_{pp}^{hJ/\psi}\sim 0.0514$ $\mu$b), calculated from a parametrization detailed in \cite{PRL116-222301}.

It is not possible to measure the individual contribution of each mechanisms of production (hadroproduction and photoproduction) in the kinematical regime of the interest. Therefore, the number of the $J/\psi$ events showed in (\ref{eq:raa}) was separated in two terms
\begin{eqnarray}
N_{AA}^{J/\psi}=\underset{\text{hadro}}{\underbrace{N_{AA}^{hJ/\psi}}}+\underset{\text{photo}}{\underbrace{N_{AA}^{\gamma J/\psi}}}.
\end{eqnarray}

Considering the central data and the hypothesis shown in the ALICE paper \cite{PRL116-222301}, it is possible to write the number of the $J/\psi$ events as a function of the average rapidity distribution as \cite{1804.09836} 
\begin{eqnarray*}
N_{AA}^{J/\psi}=\begin{cases}
1.96\times 10^6\,\,d\sigma^{\gamma}_{J/\psi}/dy, & 30\%-50\%\\
1.34\times 10^6\,\,d\sigma^{\gamma}_{J/\psi}/dy, & 50\%-70\%\\
0.96\times 10^6\,\,d\sigma^{\gamma}_{J/\psi}/dy, & 70\%-90\%
\end{cases}
\end{eqnarray*}
where $d\sigma^{\gamma}_{J/\psi}/dy$, evaluated in the $2.5<y<4.0$ rapidity range. 

\section{Theoretical Framework}\label{ipdpf}

In the ultrarelativistic limit, the differential cross section in the rapidity $y$ and impact parameter $b$, is given by \cite{PRC93-044912} 
\begin{eqnarray}
\dfrac{d^3\sigma_{AA\rightarrow AAV}}{d^2bdy} = \omega N(\omega,b)\sigma_{\gamma A\rightarrow VA}+\left(y\rightarrow -y\right)\label{pri}.
\end{eqnarray}
where $N(\omega,b)$ is a photon flux with b-dependence, $\sigma_{\gamma A\rightarrow VA}$ is the photonuclear cross section, which characterizes the photon-target interaction $\gamma A\rightarrow VA$, $\omega=\frac{1}{2}M_V\text{exp}(y)$ is the photon energy and $M_V$ is the meson mass. In the peripheral collisions ($b<2R_A$), the use of the different eletromagnetic form factor, $F(k^2)$, becomes relevant and, therefore, the following generic formula was adopted for the photon flux\cite{PPNP39-503}   
\begin{eqnarray}
N\left(\omega,b\right)=\dfrac{Z^{2}\alpha_{QED}}{\pi^{2}\omega}\left|\int_{0}^{\infty}dk_{\perp}k_{\perp}^{2}\dfrac{F\left(k^2\right)}{k^2}J_{1}\left(bk_{\perp}\right)\right|^{2}\label{eq:xx},
\end{eqnarray}
where $Z$ is the nuclear charge, $\gamma=\sqrt{s_{NN}}/(2m_{\textrm{proton}})$ is the Lorentz factor, $k_{\perp}$ is the transverse momentum of the photon and $k^{2}=\left(\omega/\gamma\right)^{2}+k_{\perp}^{2}$. A good approximation for the lead nucleus form factor is shown in \cite{PRC14-1977}, in which the Woods-Saxon distribution is rewritten as a hard sphere,
with radius $R_A$, convoluted with an Yukawa potential with range $a = 7$ fm. The Fourier transform of this convolution result in 
\begin{align}
\scalebox{0.7}{$F(k)=\dfrac{4\pi\rho_{0}}{Ak^{3}}\left[\textrm{sin }\left(kR_{A}\right)-kR_{A}\textrm{cos }\left(kR_{A}\right)\right]\left[\dfrac{1}{1+a^{2}k^{2}}\right]$},\label{eq:wsy}
\end{align}
where $A$ is the mass number, $a=0.7$ fm and $\rho_{0}=0.1385\textrm{ fm}^{-3}$.

The second factor in (\ref{pri}), $\sigma_{\gamma A\rightarrow VA}$, represents the photon-nuclei interaction and can be described in the light cone colour dipole formalism, which includes the partonic saturation phenomenon and the nuclear shadowing effects \cite{EPJC8-115}. The formalism has already been explored in the last works \cite{PRD94-094023} in pp, pA and AA collisions. In the last case, the coherent photonuclear cross section of a vector meson $V$ can be factorized as 
\begin{eqnarray}
\sigma_{(\gamma A\rightarrow VA)} = \left.\frac{d\sigma}{dt}\right|_{t=0}\int_{t_{min}}^{\infty}|F(t)|^2dt\label{sdt}
\end{eqnarray}
where the forward scattering amplitude, $d\sigma/dt|_{t=0}=|\text{Im }A(x,t=0)|^2/16\pi$, carries the dynamical information of the process and the form factor, $F(t)$, represents the dependence on the spatial characteristics of the target.

On the Eq. (\ref{sdt}) are also included the parameter $\beta=\text{tan }(\pi\lambda_{eff}/2)$, which restores the real contribution of the scattering amplitude, and $R_g^2(\lambda_{eff})=(2^{2\lambda_{eff}+3}/\sqrt{\pi})[\Gamma(\lambda_{eff}+5/2)/\Gamma(\lambda_{eff}+4)]$, which corresponds to the ratio of off-forward to forward gluon distribution (skewedness effect) relevant for heavy mesons. The parameter $\lambda_{eff}$ can be estimated from relation $\lambda_{eff}\equiv\partial\text{ln}[\text{Im }A(x,t=0)]/[\partial\text{ln}(1/x)]$. Thus, the Eq. (\ref{sdt}) is rewritten as
\begin{eqnarray}
\sigma_{(\gamma A\rightarrow VA)} &=&\frac{|\textrm{Im }A(x,t=0)|^2}{16\pi}\left(1+\beta^2(\lambda_{eff})\right)R^2_g(\lambda_{eff})\nonumber
\\
&\times&\int_{t_{min}}^{\infty}|F(t)|^2dt,\label{seg}
\end{eqnarray}
where $x=(M^2_V+Q^2)/(Q^2+2\omega\sqrt{s})$ with $Q\sim 0$ for nucleus and $t_{min}=(m_V^2/2\omega\gamma)^2$. 
In the color dipole formalism, the photon-nuclei forward scattering amplitude is factorized in the overlap between the photon and the vector meson wave functions, and in the dipole-nuclei cross section as
\begin{eqnarray}
\text{Im }A(x,t=0)=\int d^2r\int\frac{dz}{4\pi}\left(\psi^*_V\psi_{\gamma}\right)_T\sigma^{\text{nucleus}}_{\text{dip}}(x,r).
\end{eqnarray}
where $\left(\psi^*_V\psi_{\gamma}\right)_T$ is described with more detail in \cite{PRD74-074016}, and $\sigma^{\text{nucleus}}_{\text{dip}}(x,r)$ is obtained using the Glauber-Gribov picture \cite{JETP29-483,JETP30-709}, as proposed in \cite{EPJC26-35}
\begin{eqnarray}
\scalebox{0.9}{$\sigma_{\textrm{dip}}^{\textrm{nucleus}}(x,r)=2\int d^2b'\left\{1-\textrm{exp}\left[-\frac{1}{2}T_A(b')\sigma_{\textrm{dip}}^{\textrm{proton}}(x,r)\right]\right\}$}\label{nuc}.
\end{eqnarray}
In the Eq. (\ref{nuc}), $T_{A}(b)$ is the nuclear overlap function and $\sigma_{dip}$ is the dipole-nucleon cross section, which was calculated in this work using the GBW and CGC dipole models. These two models shown a good agreement with the data in the ultraperipheral regime \cite{PRD94-094023}. The application of the equations (\ref{eq:xx}) and (\ref{seg}) inside of (\ref{pri}) constitutes what we named the \textbf{scenario 1}, which produces the first results presented in Table \ref{dsigdy2}.

\section{The Effective Photon Flux}

\begin{figure}[H]
	\centering
	\scalebox{0.55}{
		\includegraphics{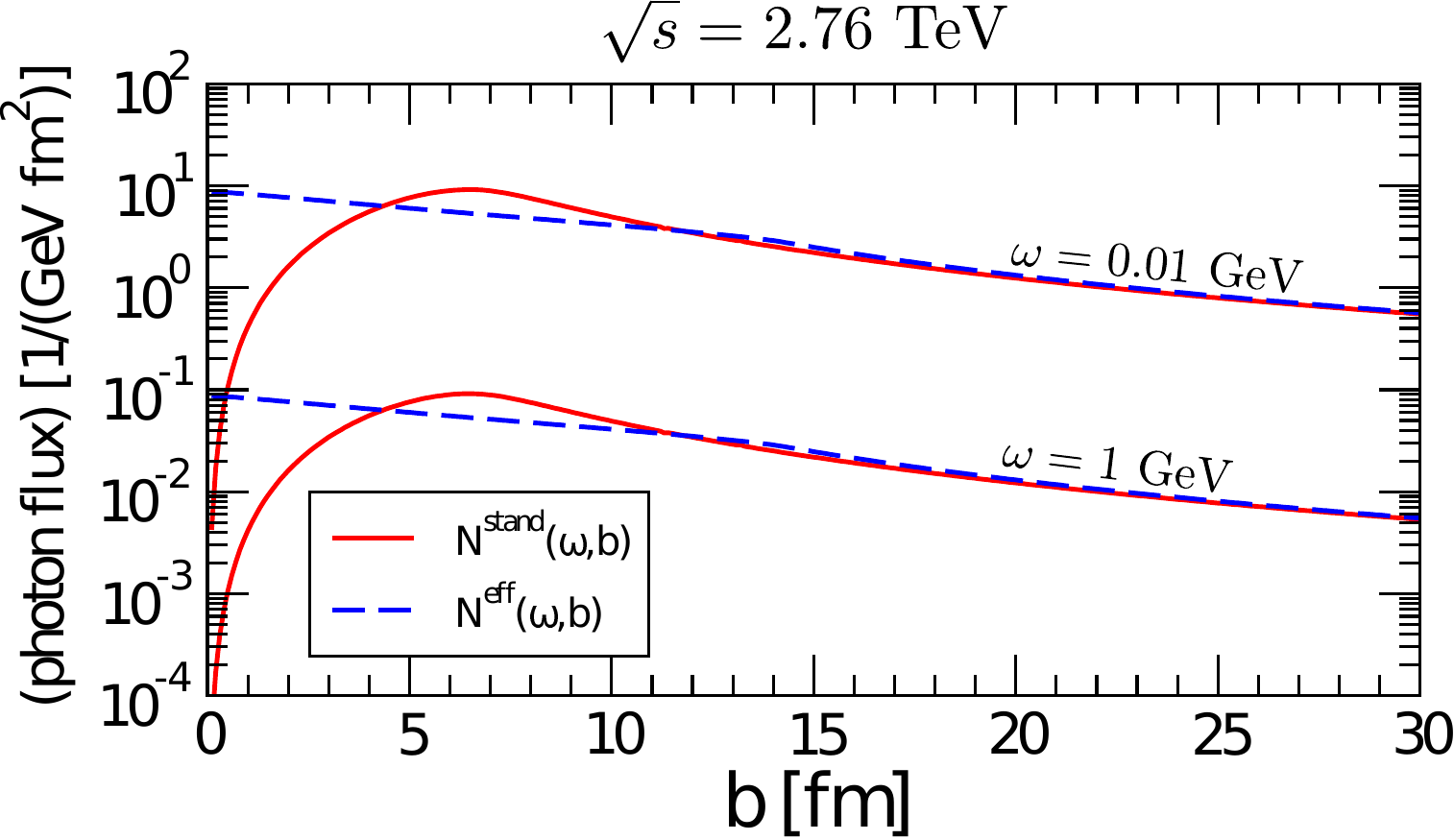}}
	\caption{Comparison between the usual and effective photon flux for the $\omega=0.01$ GeV and $\omega=1$ GeV at $\sqrt{s}=2.76$ TeV.}
	\label{fig:fig3}
\end{figure}

In order to refine our calculations, an effective photon flux was built as a function of usual photon flux (eq. \ref{eq:xx}) following a similar procedure showed in \cite{PRC93-044912} where considered two hypothesis: (1) only the photons that reach the geometrical region of the nuclear target will be considered and (2) the photons that reach the overlap region will be neglected. Then, the new photon flux can be expressed as \cite{PRD96-056014}
\begin{eqnarray}
\scalebox{0.9}{$N^{eff}(\omega,b)=\int N^{usual}(\omega,b_1)\frac{\theta(b_1-R_A)\theta(R_A-b_2)}{A_{eff}(b)}d^2b_2$}\label{Neff}
\end{eqnarray}
where the effective interaction area, $A_{eff}(b)$, is given by
\begin{align}
\scalebox{0.9}{$A_{eff}(b)=R_A^2\left[\pi-2\textrm{cos}^{-1}\left(\frac{b}{2R_A}\right)\right]+\frac{b}{2}\sqrt{4R_A^2-b^2}$},
\end{align}
In the Figure \ref{fig:fig3}, the usual photon flux is compared with the effective photon flux for energy of the photon $\omega = 0.01$ GeV and $\omega = 1$ GeV (the photon flux is dominated by photons with energy $\lesssim 0.2$ GeV). As expected, both models become similar as $b\rightarrow\infty$. In the range $4\text{ fm}\lesssim b\lesssim 11\text{ fm}$, the usual photon flux is bigger than the effective photon flux, mainly on the threshold $b\sim R_A\sim 7$. Lastly, the usual photon flux diverges considerably from the effective photon flux as $b\rightarrow 0$.

\begin{figure}[h]
	\centering
	\scalebox{0.55}{
		\includegraphics{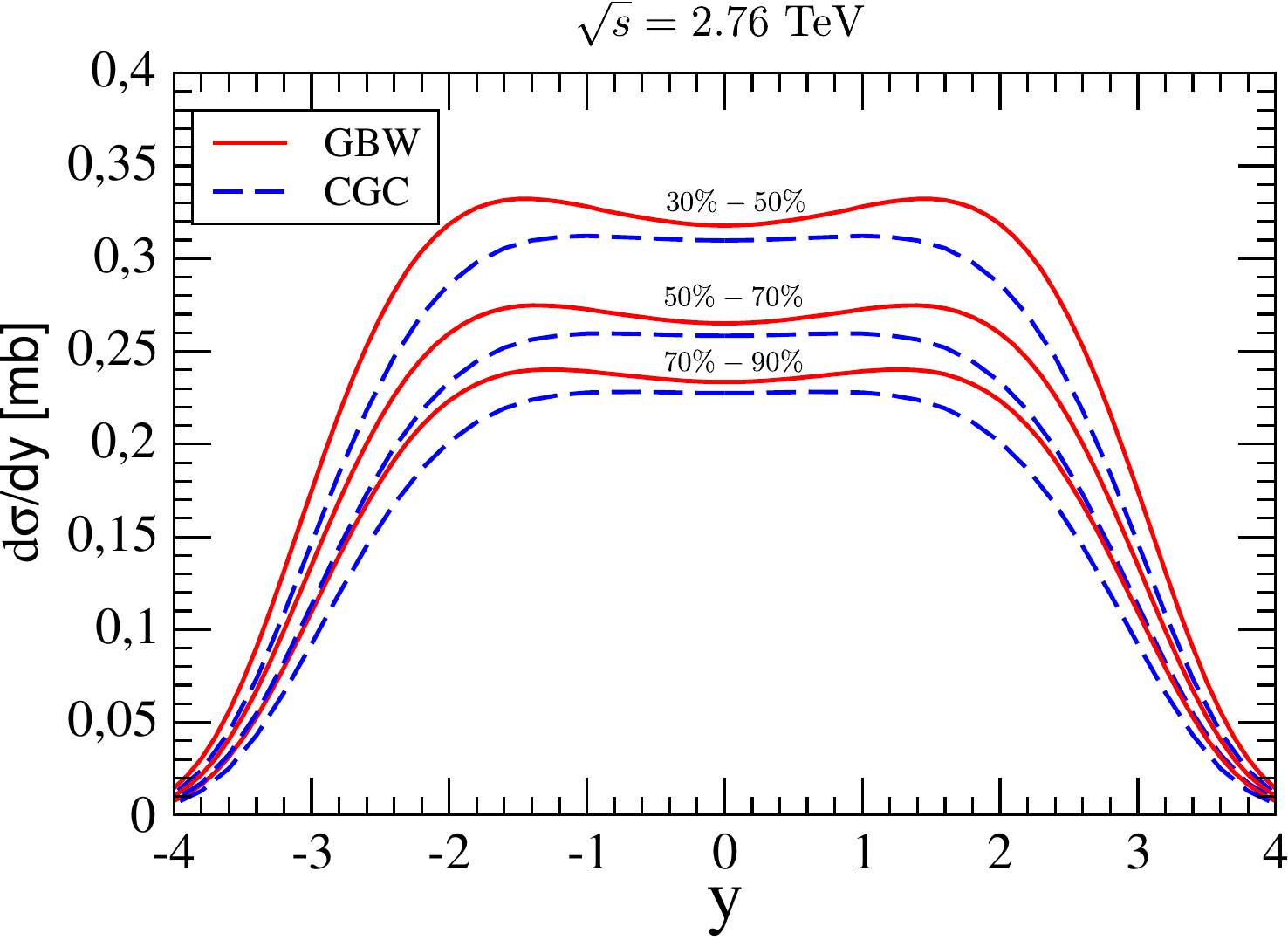}}
	\caption{Rapidity distribution for $J/\psi$ nuclear photoproduction at $\sqrt{s}=2.76$ TeV for different centrality classes using the GBW and CGC dipole models.}
	\label{psi1s}
\end{figure}

\begin{figure}[h]
	\centering
	\scalebox{0.55}{
		\includegraphics{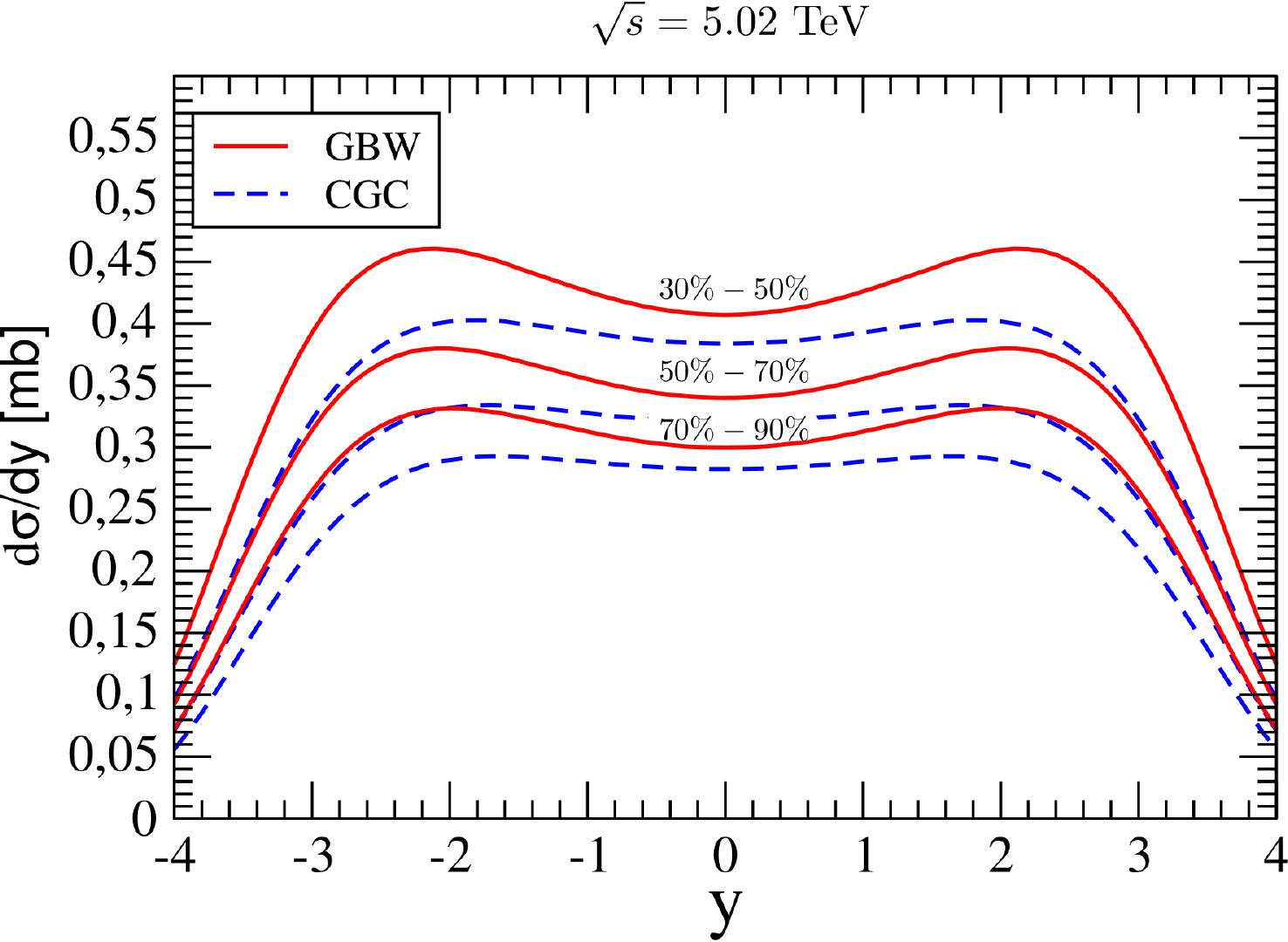}}
	\caption{Rapidity distribution for $J/\psi$ nuclear photoproduction at $\sqrt{s}=5.02$ TeV for different centrality classes using the GBW and CGC dipole models.}
	\label{psi1s5020}
\end{figure}

Considering the effective photon flux (Eq. (\ref{Neff})) and the photonuclear cross section (\ref{seg}), the rapidity distribution for $J/\psi$ photonuclear production was calculated in Pb-Pb collisions at $\sqrt{s} = 2.76$ TeV (Fig. \ref{psi1s}) and $\sqrt{s} = 5.02$ TeV (Fig. \ref{psi1s5020})for $y<|4.0|$. For each centrality class there is some difference in the $|y|\gtrsim 1.0$ range but, in general, the curves showed a similar behavior in relation to dipole models used. Moreover, analysing the different centrality classes, it was observed an increase of $\sim$ 12\% from 70\%-90\% to 50\%-70\% and of $\sim$ 13.7\% from 50\%-70\% to 30\%-50\%, for both dipole models, at $y=0$. Similarly, at $\sqrt{s}=5.02$ TeV, it was observed an increase of the $\sim$ 12\% from 70\%-90\% to 50\%-70\% and $\sim$ 13.3\% from 50\%-70\% to 30\%-50\% at $y=0$. Therefore, in the central region, the relative variation between the different centrality classes is not sensitive to the increase of the energy.      

The ratio $\frac{d\sigma^{5.02}}{dy}/\frac{d\sigma^{2.76}}{dy}$ was also investigated. An increase of approximately 30\% in the central rapidity region $|y|<1.5$ was observed for the three centrality classes analyzed. However, this ratio is of the 60\% for the same rapidity region in the UPC. Therefore, in this formalism the effective photon flux seems less sensitive with the variation of the energy in comparison with the usual photon flux.

\section{The Effective Photonuclear Cross Section}

To ensure more consistence in the application of the effective photon flux, the photonuclear cross section also needs to be limited in accordance with the geometrical constraints adopted in the construction of the effective photon flux, in order to consider only the interaction of the photon with the non-overlap region. This is achieved by applying the $\Theta\left(b_1-R_A\right)$ function into Eq. (\ref{nuc}), which produces
\begin{eqnarray}
\scalebox{1.0}{$\sigma_{\textrm{dip}}^{\textrm{nucleus}}(x,r)$}&\scalebox{0.7}{$=$}&\scalebox{0.7}{$2\int d^2b_2\Theta(b_1-R_A)$}\nonumber
\\
& \scalebox{0.7}{$\times$}
&\scalebox{0.7}{$\left\{1-\textrm{exp}\left[-\frac{1}{2}T_A(b_2)\sigma_{\textrm{dip}}^{\textrm{proton}}(x,r)\right]\right\}$}\label{sigeff}
\end{eqnarray}
where, $b_1^2=b^2+b_2^2+2bb_2\textrm{cos}(\alpha)$. Considering the effective photon flux and photonuclear cross section, the rapidity distribution was calculated and its results for the three centrality classes (scenario 3) are shown in the Table (\ref{dsigdy2}).

\section{Main Results}\label{result}

The results of the average rapidity distribution for the three scenarios described in the text are shown in the Table (\ref{dsigdy2}), where were taken into account the GBW and CGC dipole models. 
\begin{table}[H]
	\centering
	\renewcommand{\arraystretch}{1.5}
	\scalebox{0.9}{
		\begin{tabular}{|c|c|c|c|}
			\multicolumn{4}{c}{Average Rapidity Distribution: $d\sigma/dy$}\tabularnewline
			\hline
			{\color{red}GBW}/{\color{blue}CGC} & 30\%-50\% & 50\%-70\% & 70\%-90\%\tabularnewline
			\hline 
			\textbf{Scenario 1} & {\color{red}200}/{\color{blue}170} & {\color{red}100}/{\color{blue}84} & {\color{red}60}/{\color{blue}51}\tabularnewline
			\hline 
			\textbf{Scenario 2} & {\color{red}128}/{\color{blue}107} & {\color{red}98}/{\color{blue}80} & {\color{red}80}/{\color{blue}67}\tabularnewline
			\hline 
			\textbf{Scenario 3} & {\color{red}73}/{\color{blue}61} & {\color{red}78}/{\color{blue}66} & {\color{red}75}/{\color{blue}63}\tabularnewline
			\hline 
			ALICE data & {\color{purple}$73\pm44^{+26}_{-27}\pm10$} & {\color{purple}$58\pm16^{+8}_{-10}\pm8$} & {\color{purple}$59\pm11^{+7}_{-10}\pm8$}\tabularnewline
			\hline 
		\end{tabular}}
		\caption{Comparison between our results obtained from different approximations and the ALICE data \cite{PRL116-222301}.\label{dsigdy2}}
\end{table}

It is observed that in the simplest approach (scenario 1), there is a good agreement with the ALICE data in the more peripheral region where the b-dependence is more weak. For the more central regions, the scenario 2 and scenario 3 are more suitable although both overestimate the central value of the ALICE in the 70\%-90\% region. In particular, the scenario 3 one has the largest production cross section in the 50-70\% centrality class. This is due the dipole-nuclear cross section, which is b-dependent through the $T_A(b)$ function, to be more strongly suppressed in more central collisions. Consequently, the 30\%-50\% centrality class is more affected than the 50\%-70\% region. 

\begin{figure}[H] 
	\centering
	\scalebox{0.5}{
		\includegraphics{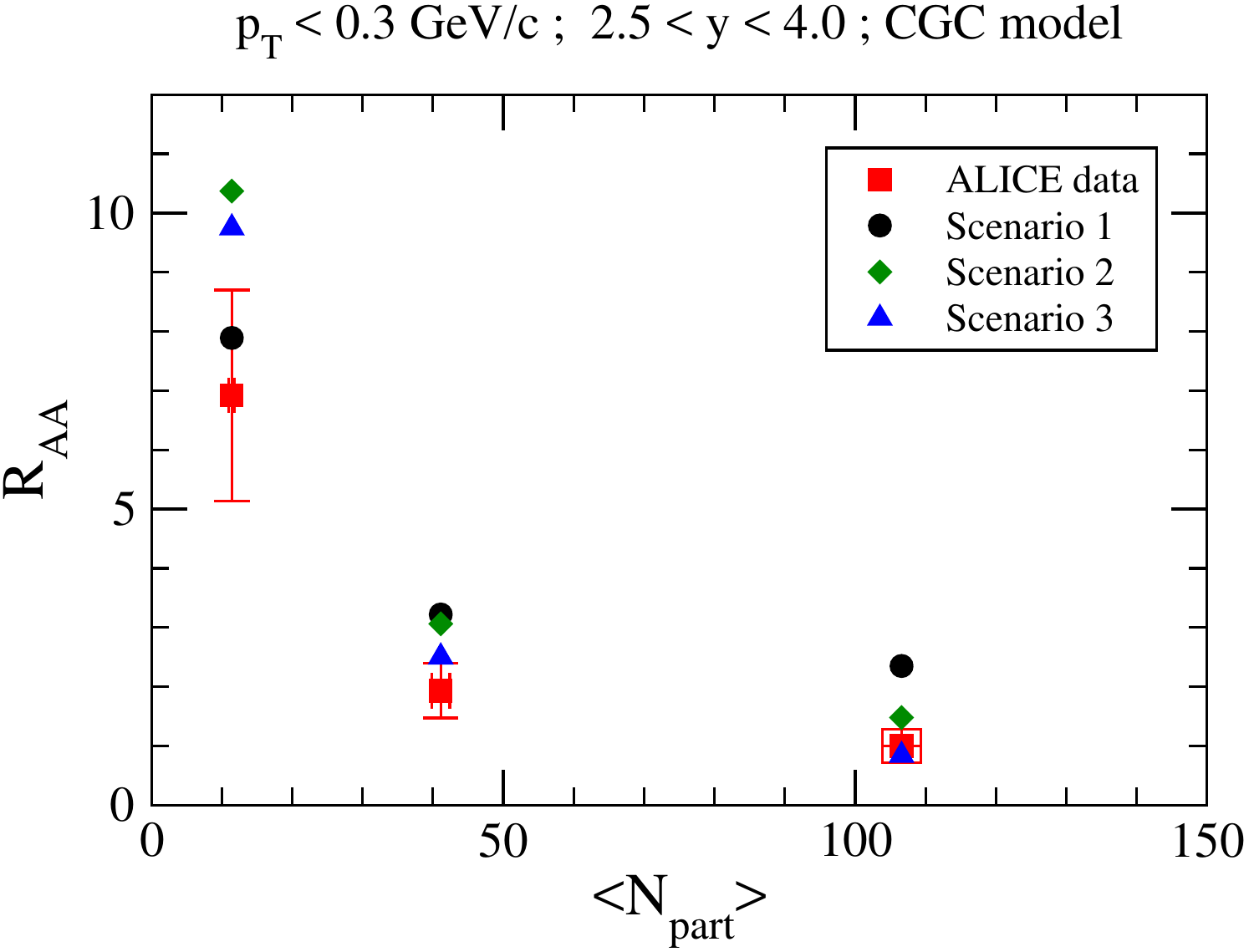}}
	\caption{Comparison of the $R_{AA}$ results with the ALICE data for the centrality classes 30\%-50\%, 50\%-70\% e 70\%-90\% \cite{PRL116-222301}.}
	\label{raa}
\end{figure}

In addition to rapidity distribution, the excess of the $J/\psi$ also was quantified by nuclear modification factor, experimentally defined by Eq. (\ref{eq:raa}), and calculated from the results presented in the Table (\ref{dsigdy2}). Adopting the CGC model, which shows slightly better results than GBW model, the $R_{AA}$ was calculated for the three scenarios investigated and its results are compared with the ALICE data, Fig. \ref{raa}. Similarly to rapidity distribution, the scenario 1 show better agreement in the more peripheral region while the scenarios 2 and 3 are more suitable for more central collisions where the b-dependence is more relevant. More details about each scenario can be found in \cite{1804.09836}.

\section{Summary}

In this work, the estimates for the rapidity distribution and nuclear modification factor were presented for the $J/\Psi$ production in the centrality classes 30\%-50\%, 50\%-70\% and 70\%-90\%. The ALICE measurements were compared with our estimates, obtained from three different approaches. In the simplest approach (scenario 1), better aggrement was obtained  with the data only in the more peripheral region, where there is a considerable uncertainty. For the more consistent approach (scenario 3), the result overestimate in the more peripheral region, however, it agrees better with the data in more central region, where the color dipole formalism is more intensely tested. Although it is not yet possible to confirm that the exclusive photoproduction is fully responsible for the $J/\psi$ excess observed in ALICE, there are indications that it produces a considerable part of the effect.

\begin{acknowledgments}
We would like to thank Dr.\hspace{1mm}Ionut Arsene for usefull discussions. This work was partially financed by the Brazilian funding agency CNPq.

\end{acknowledgments}


\end{document}